\documentclass[prl,twocolumn,showpacs,twoside,floatfix]{revtex4}
\usepackage{graphicx,amsmath,amssymb,mathptm}
\begin{document}
\title{Distribution of extremes in the fluctuations of two-dimensional equilibrium interfaces}
\author{Deok-Sun Lee}
\affiliation{Theoretische Physik, Universit\"{a}t des Saarlandes, 
  66041 Saarbr\"{u}cken, Germany}
\date{\today}
\begin{abstract}
We investigate the statistics of the maximal fluctuation 
of two-dimensional Gaussian interfaces.  Its relation to the entropic repulsion 
between rigid walls and a confined interface is 
used to derive the average maximal fluctuation $\langle m\rangle 
\sim \sqrt{2/(\pi K)}\, \ln N$ and  
the asymptotic behavior of the whole distribution 
$P(m) \sim N^2\, e^{-{\rm (const)} \, 
  N^2 e^{-\sqrt{2\pi K}\, m} - \sqrt{2\pi K} \, m}$ for $m$ finite
with $N^2$ and $K$ the interface size and tension, respectively. 
The standardized form of $P(m)$ does not depend on 
$N$ or $K$, but shows a good agreement with 
Gumbel's first asymptote distribution with 
a particular non-integer parameter.
The effects of the correlations among individual fluctuations 
on the extreme value statistics are discussed in our findings. 
\end{abstract}
\maketitle

A common nature of floods, stock market crashes, and the Internet failures 
is that they hardly occur but, once they do, have significant consequences for 
the corresponding systems. These rare but fatal events correspond to  
the appearances of extreme values in the fluctuating variables such as  
the daily discharge of a river, the stock index, and the router load,  
and therefore it has been in great demand to be able to estimate and predict 
the typical magnitude of the largest values and  
the probability of  a given extreme value~\cite{gumbel58}. 
The extreme value theory has been of great importance also in various physics 
contexts since the physics of complex systems is often governed by  
the extreme value statistics~\cite{bouchaud97}. 
Recently, Bramwell and co-workers discovered a universal distribution 
for the order parameters of a class of model systems for 
ferromagnet,  confined turbulent flow, avalanche, and  
granular media~\cite{bramwell98,bramwell00}. 
The universal distribution turns out to be consistent with 
Gumbel's first asymptote distribution that is one of the well-known 
probability distributions, to be precise, of the $n$th largest value 
among $\mathcal{N}$ independent (uncorrelated) 
random variables~\cite{gumbel58}.  Interestingly, the parameter $n$ takes 
a non-integer value and this has been considered as related to 
strong correlations in those systems~\cite{bramwell00}. While the reason 
for this universality is still debated~\cite{aji01,clusel04}, 
it renewed the interest in the extreme value statistics of 
correlated variables~\cite{antal01,dahlstedt01,gyorgyi03}. 

In this Letter we study the statistics of the maximal fluctuation 
of two-dimensional (2D)  equilibrium interfaces.
The Gaussian ensemble is considered, where 
an interface configuration $\{\phi\} = 
\{\phi(\vec{r})|\vec{r}=(x,y), -L/2\leq x,y<L/2 \}$ 
under the periodic boundary condition has the probability 
$P[\{\phi\}]=e^{-\mathcal{H}_0[\{\phi\}]}$  with 
the Gaussian Hamiltonian  
${\cal H}_0[\{\phi\}] = \int d^2r (K/2) (\nabla \phi)^2$ and $K$ the interface tension. 
The spatial average $\overline{\phi}=L^{-2}\int d^2r\, 
\phi(\vec{r})$ is set to be zero.
This Gaussian model has been used to study rough (self-affine) 
interfaces ubiquitous in nature~\cite{barabasi95}, 
the interacting spin system at low temperatures 
through the 2D XY model with $\phi$ representing spin orientations~\cite{bramwell00}, 
and so on.
We define the maximal fluctuation $m$ for 
a given interface $\{\phi\}$ as
$m = \max_{\vec{r}\in R} \{|\phi(\vec{r})|\}$,
where   
$R\equiv\{(an_x,an_y)|n_x, n_y = -N/2+1,-N/2+2,\ldots,N/2\}$ 
with $a$ the lattice constant and $N\equiv L/a$.
It is an extreme value among the individual 
fluctuations, $\phi$'s, that are correlated  
via $\mathcal{H}_0$ 
as $\langle \phi(\vec{r})\phi(\vec{r}')\rangle\sim 
\ln (L/|\vec{r}-\vec{r}'|)$~\cite{barabasi95}, and 
we study the statistics of $m$. 
The effects of such correlations have been studied 
by a renormalization group approach in the context 
of glass transition~\cite{carpentier01}, but the nature of 
the statistics of $m$ remains to be understood.
The maximal fluctuation has 
its own technological significance as well, for instance, in 
the onset of the breakdown of corroded surfaces and the occurrence of a 
short circuit by the metal surface of one electrode reaching 
the opposite one~\cite{shapir01,majumdar04}.

The exact distributions of the maximal height~\cite{shapir01,majumdar04} and 
the width square~\cite{fortin94} 
have been obtained for the 1D Gaussian interface. 
For two dimensions, however, the distribution of the maximal fluctuation 
is not known while the width-square distribution is 
well understood~\cite{racz94}.
Of our particular interest are 
the functional form of the maximal fluctuation distribution, $P(m)$,  
and its first few moments.
We use the relation of $P(m)$ to the entropic repulsion 
between an interface and rigid walls to find 
$\langle m\rangle\equiv\int dm \; m \, P(m)$ and the asymptotic behavior 
of $P(m)$, which is shown to resemble Gumbel's first asymptote 
distribution. 
The standardized distribution of the maximal fluctuation does not depend 
on any system parameter, and furthermore, Gumbel's first asymptote 
distribution with a specific parameter fits excellently  
the standardized distribution of $m$. 
These results are compared with the statistics that would appear 
without the correlations among $\phi(\vec{r})$'s 
to illuminate 
the correlation effect on the extreme value statistics.

When an interface is confined between two rigid walls, the free 
energy is increased due to the entropic repulsion against the walls. 
The maximal fluctuation distribution $P(m)$ of a free interface 
is related to this free energy  increase  with the walls 
at $\phi=m$ and $\phi=-m$, $\Delta \mathcal{F}(m)$, as 
\begin{equation}
e^{-\Delta \mathcal{F}(m)} \equiv 
\frac{{\rm Tr}\, e^{-{\cal H}_0[\{\phi\}]} \chi_m[\{\phi\}]}
{{\rm Tr}\, e^{-{\cal H}_0[\{\phi\}]}}=\int_0^{m} dm' P(m'), 
\label{eq:free}
\end{equation}
where $\chi_m[\{\phi\}]$ is $1$ for 
$\max_{\vec{r}\in R}\{|\phi(\vec{r})|\}\leq m$ 
and $0$ otherwise, and thus  
$\chi_m[\{\phi\}]=\prod_{\vec{r}\in R} \theta(m-|\phi(\vec{r})|)$ 
with $\theta(x)=1$ for $x\geq 0$ and $0$ otherwise.
We first derive $\Delta \mathcal{F}(m)$ 
and obtain $P(m)$ using Eq.~(\ref{eq:free}). 

For the soft walls restricting 
the width square $w_2=\overline{\phi(\vec{r})^2}$ to be less than $x^2$, 
the free energy increase is given as 
$\Delta \mathcal{F}(x)\sim e^{-({\rm const}) x^2}$
~\cite{mcbryan77,fisher82,bricmont86}:
The entropy reduction is inversely proportional to 
the projected area of the interface segment $\ell^2$ that is so large that 
the average fluctuation over the segment, $\sim\sqrt{\ln \ell}$,  is comparable to $x$.  
For the rigid-wall problem,  one may expect the same result 
if the average fluctuation and the maximal fluctuation 
scale as functions of the segment area in the same way, 
which is true  for 1D interface~\cite{shapir01,majumdar04}, 
but not for the 2D one. 
Bricmont {\it et al.} have shown, by means of a rigorous proof as well as 
an heuristic argument, that $\Delta\mathcal{F}(x)\sim e^{-{\rm (const)} \, x}$ 
for 2D confined Gaussian interfaces
in the thermodynamic limit $N\to\infty$~\cite{bricmont86}. 
Reviewing briefly the heuristic argument introduced therein, 
we compute   $\Delta\mathcal{F}(x)$ 
keeping the lateral size $N$ large but finite, 
which leads us to find out $\langle m\rangle$ and two distinctive 
behaviors of $P(m)$ for $m\ll \langle m\rangle$ and $m\gg \langle m\rangle$,
respectively.

One of the most prominent effects of confinement on 
an interface is  the reduction 
of the correlation length; the height fluctuations 
are hardly correlated across the interface-wall collisions.
The correlation length can be obtained in a self consistent way 
by assuming that the effective Hamiltonian 
${\cal H}_{\rm eff}(x)$ satisfying 
${\rm Tr}\, e^{-{\cal H}_{\rm eff}(x)}/{\rm Tr}\, e^{-{\cal H}_0}= 
e^{-\Delta \mathcal{F}(x)}$ with $\Delta \mathcal{F}(x)$ in 
Eq.~(\ref{eq:free}) takes the form of a massive Gaussian Hamiltonian,  
${\cal H}_{\rm eff}(x)= 
\int d^2 r [(K/2) (\nabla \phi)^2 + (\mu(x)^2/2) 
\phi(\vec{r})^2]$.
The effective mass $\mu(x)$ is the inverse of the correlation length and 
obtained self-consistently.
Differentiating Eq.~(\ref{eq:free}) with respect to $m=x$ 
and using ${\cal H}_{\rm eff}(x)$,  
one obtains an approximate expression for the derivative of the free energy increase 
as
\begin{eqnarray}
-\frac{\partial}{\partial x} 
\Delta {\cal F}(x) &=& 
\frac{{\rm Tr}\ e^{-{\cal H}_0} \int \frac{d^2 r }{a^2} \delta(x-|\phi(\vec{r})|) 
  \prod_{\vec{r}' \ne \vec{r}} \theta(x - |\phi(\vec{r}')|)}
{{\rm Tr}\ e^{-{\cal H}_{0}}\chi_x}\nonumber\\
&\simeq&\int \frac{d^2 r}{a^2}  
\langle \delta(x-|\phi(\vec{r})|) \rangle_{{\cal H}_{\rm eff}(x)} \nonumber \\
  &\simeq& 
\frac{2\, N^2}{\sqrt{2\pi W(x)^2}} e^{-x^2/(2W(x)^2)},
\label{eq:sc1}
\end{eqnarray}
where $e^{-\mathcal{H}_0} \chi_x$ is approximated by $e^{-\mathcal{H}_{\rm eff}(x)}$  and 
$\langle A\rangle_{{\mathcal H}_{\rm eff}(x)}={\rm Tr}\, A \, 
  e^{-\mathcal{H}_{\rm eff}(x)} / \, {\rm Tr}\, e^{-\mathcal{H}_{\rm eff}(x)}$. 
Also we used the relation that 
$\langle \delta(\phi-x)\rangle_{\mathcal{H}_{\rm eff}(x)}=
e^{-x^2/(2W(x)^2)}/\sqrt{2\pi W(x)^2}$ 
with the roughness $W(x)\equiv 
\sqrt{\left\langle \overline{\phi(\vec{r})^2}\right\rangle_{\mathcal{H}_{\rm eff}(x)}}$ 
given by 
\begin{equation}
W(x) =\left( 
\sum_{\vec{q}}\frac{1}{(K|\vec{q}|^2 + \mu(x)^2)L^2}\right)^{1/2}. 
\label{eq:w2x}
\end{equation}
For Eq.~(\ref{eq:w2x}), 
we considered the Fourier components $\phi_{\vec{q}} = L^{-2} 
\int d^2r\, e^{-i\vec{q}\cdot\vec{r}}\,\phi(\vec{r})$ 
for $\vec{q}=(\pi/L)(2n_x-1,2n_y-1)$, 
which are decoupled in  
$\mathcal{H}_{\rm eff}(x)$, allowing another expression for 
$-(\partial/\partial x)\Delta \mathcal{F}(x)$ as  
\begin{eqnarray}
-\frac{\partial}{\partial x} 
\Delta {\cal F}(x) &\simeq & 
\frac{\partial}{\partial x} \left(
    \ln 
    \frac{{\rm Tr}\, e^{-\mathcal{H}_{\rm eff}(x)}}{{\rm Tr}\, e^{-\mathcal{H}_0}} \right)\nonumber\\
&\simeq& -  \frac{1}{2} \left(\frac{\partial}{\partial x} \mu(x)^2\right) 
\sum_{\vec{q}} \frac{1}{K|\vec{q}|^2+\mu(x)^2}. 
\label{eq:sc2}
\end{eqnarray}

\begin{figure}
\includegraphics[width=\columnwidth]{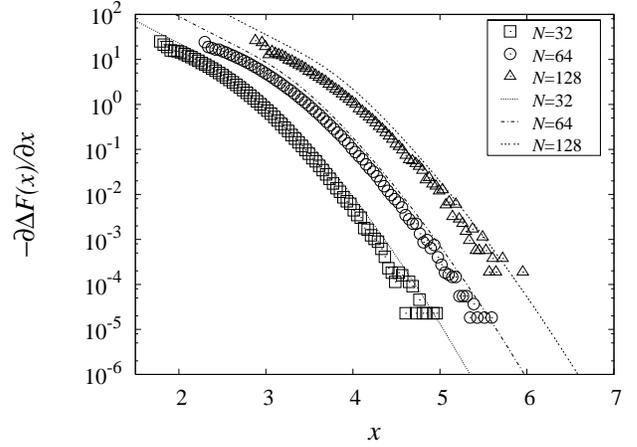}
\caption{
  Values of $-\partial \Delta\mathcal{F}(x)/\partial x$ with $K=0.92$ 
  calculated by substituting $P(m)$ from the Monte Carlo simulation 
  to Eq.~(\ref{eq:free}) [$N=32(\square)$, $64(\circ)$, $128(\triangle)$] 
  and by the self-consistent solution to Eqs.~(\ref{eq:sc1})-(\ref{eq:sc2})  
  [$N=32$ (dotted line), $64$ (dotted-dashed), $128$ (dashed)].
} 
\label{fig:rp}
\end{figure}

Equations~(\ref{eq:sc1})-(\ref{eq:sc2}) give 
a self-consistent equation for $\mu(x)$,   
which can be solved numerically with the condition $\mu(\infty) = 0$. 
A comparison of the values of $-\partial\Delta \mathcal{F}(x)/ \partial x$
from this equation and from  
a Monte Carlo simulation for free interfaces 
giving $P(m)$ and in turn, $-\partial \Delta \mathcal{F}(x)/\partial x$ 
through Eq.~(\ref{eq:free}), 
can be a test for the validity of the adopted approximation.
As shown in Fig.~\ref{fig:rp},
both are in excellent agreement 
except for a slight constant deviation that is 
presumably due to higher-order (in $\phi^2$) contributions 
to the singular factor $\theta(x-|\phi|) = \lim_{\alpha\to\infty} 
e^{-(\phi/x)^{2\alpha}}$~\cite{bricmont86}.
This justifies the use of $\mathcal{H}_{\rm eff}(x)$ 
for the study of the maximal fluctuation 
distribution of a free interface as well as 
the free energy increase of a confined interface.

\begin{figure}
\includegraphics[width=\columnwidth]{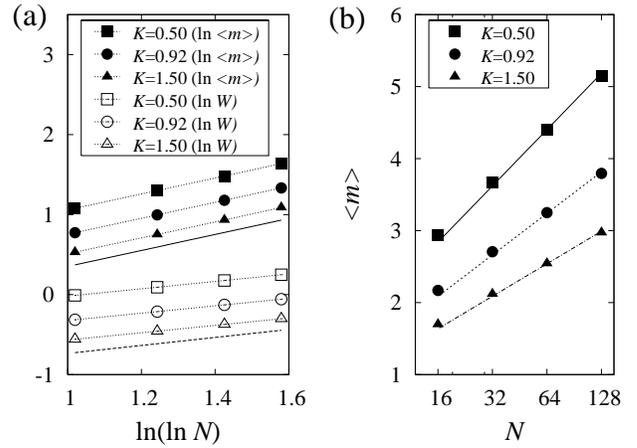}
\caption{
  (a) Plots of $\ln\langle m\rangle$ and $\ln W$ versus $\ln(\ln N)$ for 
    2D free Gaussian interfaces with $K=0.50$, $0.92$, and $1.50$.
    The upper solid line has the slope $1$ while the lower dashed line has $1/2$. 
    The dotted lines connecting data points are guide for the eye.
  (b) Semilog plot of $\langle m\rangle$ versus $N$ for the same values of $K$ as in (a).
  The lines have the slopes $\sqrt{2/(\pi K)}$, respectively.
}
\label{fig:maxave}
\end{figure}

The large-$x$ behavior of the effective mass $\mu(x)$ 
can be derived analytically.
From Eq.~(\ref{eq:w2x}), 
$W(x)^2$  is represented as 
$(2\pi K)^{-1}\ln[\mu_0/\mu(x)]$ for $\mu_0/\mu(x)\ll N$ and 
$(2\pi K)^{-1}\ln N$ for $\mu_0/\mu(x)\gg N$, where $\mu_0 = \sqrt{K}\pi/a$. 
Solving Eqs.~(\ref{eq:sc1}) and (\ref{eq:sc2}) with these values of $W(x)^2$, 
one finds that, for large $x$,  
$\mu(x)/\mu_0\sim e^{-\sqrt{\pi K/2}\, x}$ for $1\ll x\ll x_c$ and 
$\mu(x)/\mu_0\sim e^{-\pi K x^2/(2\ln N)}$ for $x\gg x_c$ 
with $x_c = \sqrt{2/(\pi K)}\ln N$.
It is natural to think that the interaction between the walls and the 
interface is negligible and that the correlation length diverges  
when the distance to the wall $x$ is much larger than $\langle m\rangle$. 
Therefore, we expect the characteristic distance $x_c$ is equal to $\langle m\rangle$, 
\begin{equation}
\langle m \rangle \simeq  
\sqrt{\frac{2}{\pi K}}\ln N.
\label{eq:maxave}
\end{equation}
This relation is in agreement with the result in Ref.~\cite{carpentier01} and 
is also verified by our Monte Carlo simulation data 
for $\langle m\rangle$ as a function of $N$ plotted in Fig.~\ref{fig:maxave}. 
The scaling of $\langle m\rangle$ is distinguished from that of 
the roughness $W\sim \sqrt{\ln N}$ [Fig.~\ref{fig:maxave}].
This is contrasted to the 1D Gaussian interface, where 
both the maximal fluctuation and the roughness 
scale as $\sqrt{N}$~\cite{shapir01,majumdar04}. 

Using the obtained functional form of $\mu(x)$, 
one can also obtain the  
free energy increase  $\Delta\mathcal{F}(x)$ and 
in turn, the maximal fluctuation distribution 
$P(m)$ through Eq.~(\ref{eq:free}): 
\begin{equation}
P(m) \sim \left\{
  \begin{array}{ll}
  N^2\, e^{-{\rm (const)}\,N^2\, e^{-\sqrt{2\pi K}\,m} - \sqrt{2\pi K} \, m} & (1\ll m\ll \langle m\rangle),\\
  N^2\, e^{-{\rm (const)}\, N^2 e^{-\frac{\pi K\, m^2}{\ln N} } - \frac{\pi K\, m^2}{\ln N}} & 
  (m\gg \langle m\rangle).
  \end{array}
  \right.
\label{eq:pdf}
\end{equation}
The asymptotic behavior of $P(m)$ for $m\ll \langle m\rangle$ in Eq.~(\ref{eq:pdf}) is consistent with Gumbel's first asymptote distribution 
that represents  
the limiting distribution of the $n$ th largest value, $m_n$,  
among $\mathcal{N}\to\infty$ independent random variables
following an exponential-type distribution~\cite{gumbel58}. 
In terms of the standardized variable $z=(m_n-\langle m_n\rangle)/\sigma_{m_n}$ 
with $\langle m_n\rangle$ and $\sigma_{m_n}$ the average and the standard 
deviation of $m_n$, Gumbel's  first asymptote 
distribution  takes 
the standardized form $g(z;n)$ given by~\cite{gumbel58,bramwell00}
$g(z;n)=\omega \  e^{-n[e^{-b(z+s)}+b(z+s)]}$,
where $b=\sqrt{\psi'(n)}$, $s=(\ln n-\psi(n))/b$, and $\omega=n^n b/\Gamma(n)$ 
with $\Gamma(x)$ the Gamma function and $\psi(x)=\partial \ln \Gamma(x)/\partial x$~\cite{bramwell00}.

\begin{figure}
\includegraphics[width=\columnwidth]{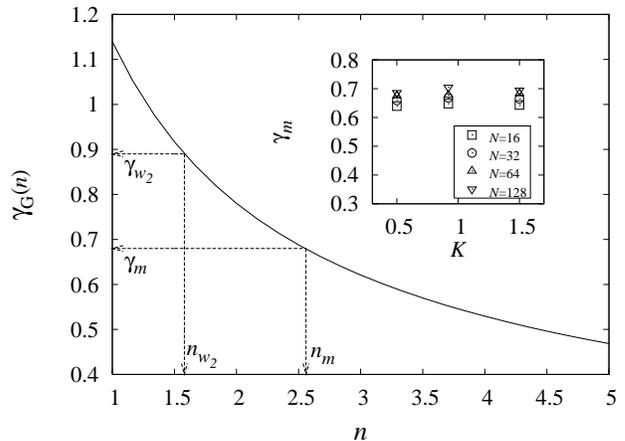}
\caption{
  Skewness of the Gumbel distribution 
  $\gamma_{\rm G}(n) \equiv \int dz\, z^3 g(z;n)$. 
  It is shown that $\gamma_{\rm G}(n)$ with $n=n_m\simeq 2.6(2)$ is 
  equal to  the skewness of $P(m)$, $\gamma_m$, which 
  is $0.68(2)$ independent of the system size $N$ or tension $K$ as 
  shown in the inset.
  The skewness of the width-square distribution, $\gamma_{w_2}$, is about $0.89$, 
  which is reproduced by $g(z;n)$  with $n=n_{w_2}\simeq 1.58$~\cite{bramwell00}.}
\label{fig:skew}
\end{figure}
\begin{figure}
\includegraphics[width=\columnwidth]{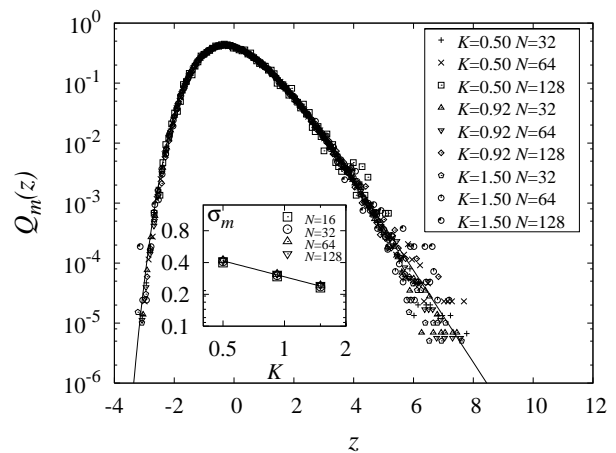}
\caption{
Standardized distribution $Q_m(z)$  for $N=32$, $64$, $128$, and 
$K=0.5$, $0.92$, $1.5$. The data points in the range 
$z\lesssim 5$ fall on 
the Gumbel distribution $g(z;n_m)$ with $n_m\simeq 2.6$ 
represented by the solid line.
Inset: The standard deviation $\sigma_m$ of the maximal fluctuation  
is fitted by $0.29(1)\, K^{-1/2}$ (solid line) without a significant $N$-dependence. 
} 
\label{fig:pdf}
\end{figure}

It is $g(z;n)$ with a non-integer value of $n$ that 
fits the universal distribution in a class of correlated 
systems~\cite{bramwell98,bramwell00}.
The 2D Gaussian interface 
belongs to this class since
the  standardized distribution of the width square 
$Q_{w_2}(z)=\sigma_{w_2} P(w_2=\langle w_2\rangle+z\,\sigma_{w_2})$ with 
$\langle w_2\rangle$ and $\sigma_{w_2}$ the average and the standard 
deviation of $w_2$, respectively,  is fitted by  
$g(z;n=n_{w_2}\simeq 1.58)$ without regard to the system size $N$ 
or tension $K$~\cite{bramwell98,bramwell00}. 
To check whether such a unique distribution also exists for  
the maximal fluctuation,  
we computed the skewness of $P(m)$,   
$\gamma_m\equiv \langle(m-\langle m\rangle)^3\rangle/\sigma_m^3$ 
 with $\sigma_m$ the standard deviation of $m$.
According to our data shown in Fig.~\ref{fig:skew}, 
$\gamma_m$ hardly varies with 
$N$ or $K$, but stays at $0.68(2)$, which  
leads us to expect a similar universality in the maximal fluctuation 
statistics to in the width-square one. 
That particular value of the skewness $\gamma_m$ is reproduced in 
Gumbel's first asymptote distribution with $n=n_m\simeq 2.6$ as 
shown in Fig.~\ref{fig:skew}, and we compare 
the standardized distribution $Q_m(z) = \sigma_m \, 
P(m=\langle m\rangle + z\, \sigma_m)$ for different values of $N$ and $K$ 
with Gumbel's first asymptote distribution 
$g(z;n_m)$ in Fig.~\ref{fig:pdf}. One can identify
the agreement of $Q_m(z)$'s and $g(z;n_m)$ as well as 
the data collapse of $Q_m(z)$'s for different $N$'s and $K$'s. 
In addition, this agreement enables one to predict that 
$\sigma_m$ is given by 
$\sigma_m=b/\sqrt{2\pi K}$ with $b=\sqrt{\psi'(n_m)}\simeq 0.69(3)$, 
which is confirmed numerically in Fig.~\ref{fig:pdf}.

Now let us look into the effects of the correlations among 
$\phi(\vec{r})$'s on $\langle m\rangle$  
in Eq.~(\ref{eq:maxave}) and  on $P(m)$ in Eq.~(\ref{eq:pdf}).
For comparison, we consider the statistics of $m$ that 
would emerge with {\it no} correlation for 
a Gaussian distribution of $\mathcal{N}$ individual fluctuations 
$p(\phi)=e^{-\phi^2/(2W^2)}/\sqrt{2\pi W^2}$.
In this case we could apply the following relation to obtain 
$\langle m\rangle$: $\int_{-\langle m\rangle}^{\langle m\rangle} d\phi \, 
p(\phi) \simeq (\mathcal{N}-1)/\mathcal{N}$~\cite{gumbel58}.
This relation yields 
$\langle m\rangle\sim W\sqrt{\ln (\mathcal{N}/W)}$~\cite{shapir01}. 
The $\sqrt{\ln \mathcal{N}}$-scaling of 
$\langle m\rangle$ is thus generic when the roughness $W$ is finite. 
A different scaling  
of $\langle m\rangle$ shows up with a diverging roughness. For instance, 
$\langle m\rangle\sim \sqrt{N\ln N}$ in 1D Gaussian interfaces 
where $\mathcal{N}=N$ and $W\sim \sqrt{N}$ and 
$\langle m\rangle\sim \ln N$ in 2D ones where $\mathcal{N}=N^2$ 
and $W\sim\sqrt{\ln N}$. 
While the $\sqrt{N\ln N}$ behavior deviates from the 
true behavior of $\langle m\rangle$ in the 1D Gaussian 
interface~\cite{shapir01,majumdar04},
the $\ln N$ behavior in two dimensions is consistent with Eq.~(\ref{eq:maxave}). 
Without the correlations, the distribution of the maximal fluctuation 
could also be obtained by~\cite{gumbel58}
$P(m)\simeq 2\,\mathcal{N}\,\left(\int_{-m}^m d\phi \, p(\phi)\right)^{\mathcal{N}-1} 
\, p(m)$. This gives 
$P(m)\sim \mathcal{N}\, e^{-{\rm (const)} \mathcal{N} e^{-m^2/(2W^2)} - 
  m^2/(2W^2)}$ as $P(m)$ for $m\gg \langle m\rangle$, but  
is not capable of reproducing the behavior of $P(m)$ for $m$ finite in Eq.~(\ref{eq:pdf}), 
which demonstrates the essential role of the correlations.
The extreme statistics without correlations has been found 
in small-world networks where the 
correlation length is finite due to random links~\cite{guclu04}. 
The correlation length becomes finite also in the presence 
of external field~\cite{portelli01}.  
$P(m)$  is more skewed, that is, 
the skewness is larger, with larger external field~\cite{leenote}. 
On the contrary,  the width-square distribution
recovers its Gaussian form as it loses the 
correlations~\cite{portelli01}.

Our results demonstrate the availability of Gumbel's first 
asymptote distribution with a non-integer parameter $n$ 
for the maximal fluctuation distribution of the 2D Gaussian interface.
The value of $n$ can be different 
according to the type of the correlations:
It takes a value close to $\pi/2$ in the case of 
the maximal avalanche distribution in the 
Sneppen depinning model~\cite{dahlstedt01}, different from 
$2.6(2)$ found for the maximal fluctuation distribution in this work.
We remark that the value of $n_m$ is larger than $n_{w_2}$, 
that is, $\gamma_{w_2}>\gamma_m$, implying that the width-square 
distribution is more non-Gaussian than the maximal fluctuation distribution. 

In summary, we investigated the statistics of the maximal fluctuation 
in the 2D Gaussian interface. Because of the correlations among 
individual fluctuations, the average maximal fluctuation  
as well as the whole distribution exhibit distinct behaviors 
from those of uncorrelated fluctuations.
We also identified the appropriateness of Gumbel's first asymptote 
distribution with a non-integer parameter for the maximal fluctuation statistics, 
which suggests its applicability 
to the extreme value statistics in a class of correlated systems 
with the parameter depending on the correlations. 

We acknowledge valuable discussions with Marcel den Nijs and 
Doochul Kim, and helpful comments on the correlation effect  
from Yonathan Shapir.  
This work is supported by  the Deutsche Forschungsgemeinschaft.



\begin{thebibliography}{99}
\bibitem{gumbel58}
E.J. Gumbel, {\it Statistics of Extremes} (Columbia University Press, 
New York, 1958).
\bibitem{bouchaud97}
J.P.~Bouchaud and M.~M\'{e}zard, J.~Phys.~A {\bf 30}, 7997 (1997). 
\bibitem{bramwell98}
S.T.~Bramwell, P.C.W. Holdsworth, and J.-F. Pinton, Nature {\bf 396}, 
  552 (1998).
\bibitem{bramwell00}
S.T. Bramwell {\it et al.}, Phys. Rev. Lett. {\bf 84}, 3744 (2000); 
Phys. Rev. E {\bf 63}, 041106 (2001).
\bibitem{aji01}
V.~Aji and N.~Goldenfeld, Phys.~Rev.~Lett. {\bf 86}, 1007 (2001).
\bibitem{clusel04}
M.~Clusel, J.-Y.~Fortin, and P.C.W.~Holdsworth, Phys.~Rev.~E {\bf 70}, 
046112 (2004).
\bibitem{antal01}
T.~Antal, M.~Droz, G.~Gy\"{o}rgyi, and Z.~R\'{a}cz, 
Phys.~Rev.~Lett. {\bf 87}, 240601 (2001); 
Phys.~Rev. E {\bf 65}, 046140 (2002).
\bibitem{dahlstedt01}
K.~Dahlstedt and H.J.~Jensen, J.~Phys.~A {\bf 34}, 11193 (2001).
\bibitem{gyorgyi03}
G.~Gy\"{o}rgyi, P.C.W.~Holdsworth, B.~Portelli, and 
Z.~R\'{a}cz, Phys.~Rev.~E {\bf 68}, 056116 (2003).
\bibitem{barabasi95}
A.-L.~Barab\'{a}si and H.E.~Stanley, {\it Fractal Concepts in Surface
Growth}~(Cambridge Universality Press, Cambridge, 1995). 
\bibitem{carpentier01}
D.~Carpentier and P.~Le Doussal, Phys.~Rev.~E {\bf 63}, 026110 (2001).
\bibitem{shapir01}
S.~Raychaudhuri, M.~Cranston, C.~Przybyla, and Y.~Shapir, Phys. Rev. Lett. {\bf 87}, 136101 (2001).
\bibitem{majumdar04}
S.N.~Majumdar and A.~Comtet, Phys.~Rev.~Lett. {\bf 92}, 225501 (2004).
\bibitem{fortin94}
G.~Foltin, K.~Oerding, Z.~R\'{a}cz, R.L.~Workman, and R.K.P.~Zia, 
Phys.~Rev.~E {\bf 50}, R639 (1994).
\bibitem{racz94}
Z.~R\'{a}cz and M.~Plischke, Phys.~Rev.~E {\bf 50}, 3530 (1994).
\bibitem{fisher82}
M.E. Fisher and D.S. Fisher, Phys. Rev. B {\bf 25}, 3192 (1982).
\bibitem{mcbryan77}
O.~McBryan and T.~Spencer, Commun.~Math.~Phys. {\bf 53}, 299 (1977).
\bibitem{bricmont86}
J. Bricmont, A. El Mellouki, and J. Fr\"{o}hlich, J. Stat. Phys. {\bf 42}, 743 (1986).
\bibitem{guclu04}
H.~Guclu and G.~Korniss, Phys.~Rev.~E {\bf 69}, 065104(R) (2004).
\bibitem{portelli01}
B.~Portelli, P.C.W.~Holdsworth, M.~Sellitto, and S.T.~Bramwell, 
  Phys.~Rev. E {\bf 64}, 036111 (2001).
\bibitem{leenote}
D.-S. Lee (unpublished).
\end{thebibliography}
\end{document}